\documentclass[journal,comsoc]{IEEEtran}
\usepackage[table]{xcolor}

\usepackage[T1]{fontenc}

\ifCLASSINFOpdf
\else
\fi

\usepackage{multirow}

\usepackage{placeins} 
\usepackage{afterpage} 

\usepackage{amsmath}
\usepackage{verbatim}
\usepackage{bm}
\usepackage[cmintegrals]{newtxmath}

\usepackage{balance}
\usepackage[caption=false]{subfig}
\usepackage{graphicx}
\usepackage{subfig}
\usepackage{float}
\usepackage{url}
\usepackage{tabularx}
\usepackage{tikz}
\usepackage[hidelinks]{hyperref}
\usepackage{adjustbox}
\usepackage{booktabs}
\usepackage{rotating}
\usepackage{slashbox}
\usepackage{xcolor}
\usepackage{color}
\usepackage{colortbl}
\usepackage{multirow}
\usepackage[misc,geometry]{ifsym}
\usepackage{dirtytalk}
\usepackage[normalem]{ulem}
\usepackage{color}

\usepackage{svg}
\usepackage{amsmath} 
\usepackage{physics} 
\usepackage{blindtext}
\usepackage[ruled,vlined]{algorithm2e}
\newcommand\Tstrut{\rule{0pt}{2.6ex}}       
\newcommand\Bstrut{\rule[-1.5ex]{0pt}{0pt}} 
\newcommand{\TBstrut}{\Tstrut\Bstrut} 
\definecolor{grey}{RGB}{197,197,197}
\newcommand{\orcid}[1]{\href{https://orcid.org/#1}{\includesvg[width=10pt]{orcid}}}
\usepackage{color, colortbl}
\definecolor{grey}{gray}{0.9}
\usepackage{cite}
\usepackage{lineno}
\usepackage[bottom]{footmisc}
\colorlet{mygreen}{green!60!gray}

\usepackage{tablefootnote}
\usepackage{threeparttable}
\begin{document}

\title{Unscrambling the Rectification of Adversarial Attacks Transferability across Computer Networks} 

\author{
Ehsan~Nowroozi,~\IEEEmembership{Senior Member,~IEEE,} ~Samaneh~Ghelichkhani,~\IEEEmembership{Member,~IEEE},  Imran Haider ,~\IEEEmembership{Member,~IEEE,} and Ali Dehghantanha,~\IEEEmembership{Senior Member,~IEEE,}
\IEEEcompsocitemizethanks{
\IEEEcompsocthanksitem Ehsan Nowroozi is with the Centre for Secure Information Technologies (CSIT), Queen's University Belfast, Northern Ireland, United Kingdom (e-mail: e.nowroozi@qub.ac.uk).
\IEEEcompsocthanksitem Samaneh Ghelichkhani is with the University of Leeds, Faculty of Engineering and Physical Sciences Master (Computing), Master in Advanced Computer Science, United Kingdom.
 (e-mail: samanehghelichkhani@gmail.com)
\IEEEcompsocthanksitem Imran Haider is with the Department of Natural Engineering and Sciences, Bahcesehir University (BAU), Istanbul, Turkey (e-mail: imran.haider@bahcesehir.edu.tr)
\IEEEcompsocthanksitem Ali Dehghantanha Cyber Science Lab, Canada Cyber Foundry, University of Guelph, Canada, (Email: adehghan@uoguelph.ca)
}
}
\maketitle

\begin{abstract}


Convolutional neural networks (CNNs) models play a vital role in achieving state-of-the-art performances in various technological fields. CNNs are not limited to Natural Language Processing (NLP) or Computer Vision (CV) but also have substantial applications in other technological domains, particularly in cybersecurity. The reliability of CNN's models can be compromised because of their susceptibility to adversarial attacks, which can be generated effortlessly, easily applied, and transferred in real-world scenarios.

In this paper, we present a novel and comprehensive method to improve the strength of attacks and assess the transferability of adversarial examples in CNNs when such strength changes, as well as whether the transferability property issue exists in computer network applications. In the context of our study, we initially examined six distinct modes of attack: the Carlini and Wagner (C\&W), Fast Gradient Sign Method (FGSM), Iterative Fast Gradient Sign Method (I-FGSM), Jacobian-based Saliency Map (JSMA), Limited-memory Broyden fletcher Goldfarb Shanno (L-BFGS), and Projected Gradient Descent (PGD) attack. We applied these attack techniques on two popular datasets: the CIC and UNSW datasets. The outcomes of our experiment demonstrate that an improvement in transferability occurs in the targeted scenarios for FGSM, JSMA, LBFGS, and other attacks. Our findings further indicate that the threats to security posed by adversarial examples, even in computer network applications, necessitate the development of novel defense mechanisms to enhance the security of DL-based techniques.
  
\end{abstract}


\section{Introduction}

\IEEEPARstart{I}{n} the modern era of artificial intelligence (AI), and deep neural networks (DNNs) are delivering top-notch solutions to the arduous real-time issues of society~\cite{Dong2021AApplications}. 
Currently, the cutting-edge solutions of DNNs have accomplished nearly human-paralleled proficiency. CNNs are a ubiquitous class of neural network architectures that are favored for their rampant usage.  
CNNs are gaining traction in several applications. These applications incorporate discerning menaces into a network \cite{Xu2021}. 
Antecedent research has delved into how Graph Neural Networks (GNNs) can cope with adversarial attacks \cite{Daniel2018}. The core predicament in this study is that when only some details about the graph are available or the network is limited to small changes, the accuracy of classification unremittingly dwindles.  In addition, these attacks can work with different node-categorization methods. Consequently, the CNN models transcend the GNNs concerning group labeling exactitude. 
As a result, GNNs fall short of the desirable classification performance benchmarks in comparison to the CNN model.

In contrast, avant-garde investigations have demonstrated that specific instances can be computed with the intention of deceiving machine-learning (ML) models. 
To provide further explication on this subject, to produce these malicious samples, creators implemented the Papernot et al.~\cite{Papernot2016TheSettings} algorithm to craft malicious instances. 

For a comprehensive explanation in ~\cite{Papernot2016TheSettings}, the crafting of such instances is commonly known as adversarial examples and creates actual security threats. A tangible circumstance can be the incorrect tagging of traffic signals. 
The widespread recognition and versatile implementation of CNN applications have enabled them to be the primary focus for the implementation of adversarial attacks. The intuition behind adversarial examples can be attributed to the groundbreaking research of Szegedy et al. \cite{Szegedy2013IntriguingNetworks}. 

Given the natural characteristics of ML architectures, the construction of adversarial instances tends to be utilized in ML techniques, which can also affect most DNNs. In a black-box scenario, two models with independent network structures trained with separate training sets can misclassify the same adversarial sample.  
In this respect, the attacker trains the source network (SN) and generates a conflicting test case, and then \textit{ transfers } this adversary example to the target’s network (TN), although the attacker has limited intelligence about the target's model \cite{PapernotTransferabilitySamples}. This significant characteristic is referred to as \textit{transferability}, and this property may reduce CNN's performance in diverse domains. 
Considering the architecture, the transferability characteristic can be established between the source and target models owing to the correspondence between them. With regard to test accuracy, high transferability was exhibited by the models with excellent accuracy scores. Nevertheless, transferability declines among the source models with extended capacity, which means models with a large number of hyperparameters.\\
\indent The motive of our research is to determine the extent to improve the transferability of adversarial attacks. 
Through a research lens, it is crucial to address this question because defense protocols for improved attack transferability would become considerably more challenging. In addition, if the attacker has full authorization for the SN, then the security of the TN is inevitable. The investigative proof given by Tram\`er et al.~\cite{TramerTheExamples} illustrates that the probability of transferability of manipulated samples between the SN and TN is very high in larger adversarial subspaces in the overlapping section of both models. 

\indent In this study, we only focused on two examples of CNN networks, which have recently found applications in several domains such as CV \cite{Sirinukunwattana2016LocalityImages}
, and Adversarial Machine Learning (Adv-ML) \cite{GoodfellowEXPLAININGEXAMPLES, Szegedy2013IntriguingNetworks}.
Therefore, our primary focus is to improve the probability of attack transferability. In other words, we ensure the transferability of an optimized adversarial offensive against the source network by considering a totally unknown TN. 
We performed seven distinct attacks on the two most commonly used datasets, the CIC\cite{CIC-Dataset} and UNSW\cite{UNSW-Dataset}. Our study demonstrates that generated adversarial examples are often transferable between deep network architectures.  When an attacker considers higher levels of distortion, our research findings demonstrate that controlling the strength of an attack is a viable method for boosting transferability. 
%
\subsection{Contributions}
The following is the synopsis of our contributions:
\begin{itemize}

    \item We used two models, QUB1 and QUB2, and each model comprised a different number of parameters, which failed against test cases when we performed different adversarial attacks on them. These models provide a fundamental architecture for the simulation and analysis of real-world attackers. We altered the parameters that a malicious attacker can manipulate to test the resilience of the network architectures with respect to the improvement in attack transferability within different ML systems.  

    \item Instead of the datasets used in the previous research in \cite{Nowroozi2022Dem}, we used the two most recent datasets, namely, CIC \cite{CIC-Dataset} and UNSW \cite{UNSW-Dataset} due to popularity in different domains. These datasets are CIC and UNSW and are in the form of a CSV file that we obtained after a process of many steps that we mentioned in the dataset section with details.
    
    \item To demonstrate the improvements in transferability, we performed seven distinct adversarial attacks with varying parameters. These types of attacks are C\&W, FGSM, I-FGSM, JSMA, L-BFGS, DeepFool, and PGD. 
    
    \item 
    We carried out an in-depth and novel study to improve attack transferability issues (see Section \ref{Imp_Att}) and determine the effectiveness of our approach in different mismatch scenarios in terms of datasets, architectures, or both, which we refer to as half- or complete-mismatch scenarios. Additionally, we conducted a rigorous study to determine the effects of inadequate hostile intelligence and commands on the flexibility of ML models against \textit{exploratory attacks} during the test time, and we excluded causative attacks during the training phase.

\end{itemize}

\subsection{Organization}

The rest of our paper is organized as follows. Section \ref{Related} provides an overview of related studies that explore transferability properties. Section \ref{Methodology} provides background and preliminary details on the datasets, the experimental setting, and the empirical study employed in our study. The outcomes of the experiment are provided in the \ref{Results} section, followed by an explanation of the evaluation of transferability. In the end, in \ref{Conclusions} section, we summarize our findings and propose proposals for further study.

\section{Related Work}
\label{Related}

Early research on adversarial instances showed that adversarial attacks can be transferred to other networks \cite{GoodfellowEXPLAININGEXAMPLES,Szegedy2013IntriguingNetworks, PapernotTransferabilitySamples}. 
 

Within CV workflows, malicious attacks on DL models have become a major challenge for the AI community. Therefore, the transferability of distorted examples has been studied using CV techniques in\cite{Jia2019EnhancingReduction}. The adversarial samples were suggested by this study to improve the multi-task transferability in classification, OCR, object recognition models, and the detection of explicit content.  Suciu et al. \cite{SuciuWhenAttacksb} created a blueprint to assess practical adversarial samples for image categorization, Android virus scanning, Twitter-based attack recognition, and threat anticipation, and subsequently recommended evasion (also referred to as exploratory) and causative attacks for AI systems. They demonstrated that if broad transferability is not considered, the robustness of a recent evasion attack is diminished. The transferability of such attacks was studied by the authors, who provided broad aspects that impact this characteristic. In the same type of study, the authors in \cite{DemontisWhyAttacks} explored the transferability of adversarial attack scenarios (e.g., exploratory, and causative attacks) on three datasets associated with various applications, including malicious applications for Android, face recognition, and OCR for handwritten digits. 

In this study, the authors argue that the transferability property relies on three factors: target model complexity, loss function, and gradient synchronization between the substitute (resembling model) and target models. Nonetheless, the metrics under consideration have not been extensively validated in security-focused scenarios and can limit the scope of the threat model. 

The authors of \cite{Nowroozi2022Dem} studied three transferability conditions and concluded that transferability occurs under just a few specific scenarios, but not often. In that study, it was shown that JSMA and I-FGSM adversarial attacks have higher transferability in cross-model scenarios than in adversarial scenarios (e.g., cross-training). One possible clarification is that the targeted model is most likely to overfit when both attacks impact particular features. As for the datasets used for training a certain task, it should be noted that transferability is not symmetric. 


\section{Evaluation Framework}
\label{Methodology}

\begin{table}[!t]
\centering
\caption{Summary of the networks considered with the associated datasets. \label{Networks_Scenarioss}}
\begin{tabular}{|c|c|c|}
\hline
\backslashbox{Networks}{Datasets} & \textbf{CIC} & \textbf{UNSW}\TBstrut\\ \hline

\textbf{QUB1} & $N_{QUB1}^{CIC}$ &  $N_{QUB1}^{UNSW}$\TBstrut\\\hline

\textbf{QUB2} & $N_{QUB2}^{CIC}$ & $N_{QUB2}^{UNSW}$\TBstrut\\\hline

\end{tabular}
\end{table}

We studied two network architectures, \textit{QUB1} and \textit{QUB2}, which were already trained with two popular CIC and UNSW datasets and seven adversarial attacks, to investigate the transferability issue of adversarial attacks on CNN-based networks when we strengthened the degree of attack capability according to our technique. Considering the datasets and network structures listed in Table~\ref{Networks_Scenarioss}, we developed four distinct types of network scenarios. Following that, we will demonstrate the different structural elements of the networks, as well as the evaluation carried out in this study.
%

In DL, an adversary can modify the amount used for training and testing based on the information he knows. According to the literature that exists, the adversary may conduct adversarial attacks in three modes: white-box, gray-box, and black-box attack scenarios\cite{Nowroozi2022Dem}. The adversary possesses Perfect Knowledge (PK) about the TN in a \textit{white-box} setting using a particular classification for adversarial instances. The adversary has Limited Knowledge (LK) of the TN in a \textit{gray-box} scenario. The adversary does not have knowledge of the internal information of the TN in a \textit{black-box} setup, which is a more viable and complex case than the other scenarios. Consequently, the attacker employs recurring inquiries to gather such sensitive data. Therefore, given the viability of the black-box scenario in applications, in reality, we are motivated to examine it. Thus, we provide a novel strategy for boosting the strength of attack methods when attacks move deeper inside a decision boundary while minimizing distortion.
As stated earlier in ML, an adversary can change the quantity of data employed for training and testing according to its capacity. The adversary’s ability is constrained to the manipulation of only testing data in the setting of an \textit{exploratory attack} because the modifications to the training data are restricted. In contrast, in the \textit{causative attack} scenario, there is a potential risk of attack during the training stage; such types of attacks are also termed poisoning or backdoor attacks. Exploratory attacks have recently been associated with a considerable number of DL-based attacks, thereby demonstrating the importance and efficacy of these attacks. The key objective of our study is to explore an exploratory attack setting in which the adversary's capability base of knowledge is restricted to alterations to the testing data. Whereas causative scenarios (e.g., backdoors or poisoning) can disrupt the training phase by altering training examples, which we did not consider in this study. 
Many attack scenarios developed in ML and DL, have been investigated based on exploratory attack settings. 
These test-time attacks manipulate the target sample and push it beyond the model's decision boundary without impacting the training process or the decision border itself. 
%
%

\subsection{Datasets}
To evaluate the impact of adversary capability on improving transferability, we explored a wide range of real-world datasets, including both malicious and neutral examples. Therefore, we used two datasets that are highly related to our research domain. We used popular datasets, which have become part of the most recent research, to obtain high scores for training \cite{Nowroozi2023EnsembleClass}. In our study, SN and TN training was performed on the CIC\cite{CIC-Dataset} and UNSW\cite{UNSW-Dataset} datasets. We collected these datasets in the form of ‘pcap’ files, and after this, we used NFStream \cite{aouini2022nfstream} and other such tools to convert the files from pcap files to a usable format for our models. NFStream is a multi-platform Python framework that is used in Network Forensics, security analysis, network traffic monitoring, and performance analysis.

\subsection{Network Architectures}

The network formed on the CNN architecture included an input layer, convolutional layers, max-pooling layers, and a single fully connected layer. Therefore, we set up \textit{QUB1} and \textit{QUB2} models that were trained with different types of datasets (see Table \ref{Networks_Scenarioss}) and considered them to be SN, TN, and interchangeability. The setup details of the \textit{QUB1} and \textit{QUB2} models are outlined as follows.
\hfill
\subsubsection{\textbf{QUB1 Network}}
The framework of \textit{QUB1} is a deep DNN that is made with nine convolutional layers, two max-pooling layers, and one fully connected layer. Recently, this network has been employed in different domains owing to its popularity and performance, including in \cite{Nowroozi2022Dem}. 
We specified a kernel size of $3 \times 3$ and a stride of one for each convolutional layer. In addition, we considered max-pooling with a kernel size of $2 \times 2$ and a stride of 2. It is important to note that by halving the total number of feature mappings in the last convolutional layer, we reduce the number of parameters in this layer. 
%

\subsubsection{\textbf{QUB2 Network}}
In this architecture, the number of layers is less than \textit{QUB1} in comparison with \textit{QUB2}; therefore, we refer to it as a shallow model. This network was also considered recently in \cite{Nowroozi2022Dem}, owing to its popularity and performance. The \textit{QUB2} network is shallower than the \textit{QUB1} network because it has three CNN layers. However, we maintained identical configurations as the \textit{QUB1} network such that there is only one max-pooling layer and one fully connected layer, and we have three convolutional layers in this network.

\subsection{Experimental Setup} 

To build our \textit{QUB1} networks, since the \textit{QUB1} is a deep network, we considered a large number of samples for the training process, and we believe that this number of samples is sufficient for the network to generalize well. In this case, we considered 60000 patches for training, 19000 for validation, and 15000 for the test set to build the models known as $N_{QUB1}^{CIC}$ and $N_{QUB1}^{UNSW}$ per class. The patch size is set to $64 \times 64$ for all scenarios. To build our shallow network \textit{QUB2}, which is associated with $N_{QUB2}^{CIC}$ and $N_{QUB2}^{UNSW}$, we utilized 21000 patches for training, 6000 patches for validation, and 5000 patches for testing. In general, we determined that the given number of features is sufficient for our experiment to meet our research expectations. 
We developed CNN networks with Keras \cite{chollet2015} with backend Tensorflow in the Python language. We performed our analysis using hardware specification MSI GF65. We shared all the implementation codes that are publicly available in the GitHub repository \cite{ehsannowroozi_Improve_Trans}. Regarding networks, we considered 10 epochs for training, using the Adam optimizer with a learning rate of $1e^{-4}$ and a momentum of 0.99. The batch size for training and validation is set to 32, and for testing is to 100. 

\subsection{Empirical Study}

During the experiments, we employed seven unique adversarial attacks on four networks  (see Table ~\ref{Networks_Scenarioss}) to demonstrate the improvement in transference over a wide scope. With concrete examples, for every two separate classes of networks, named SN and TN, we examined the improvement of transferability from SN to TN against different adversarial attacks. 
We addressed the improvement in transferability issues between the SN and TN for each of the two different types of networks by considering various adversarial attacks. In correspondence with \cite{PapernotTransferabilitySamples}, we addressed the transferability issue by evaluating three distinct conditions to analyze the proposed approach employed to improve the transferability of adversarial attacks between SN and TN. These possibilities correspond to the mismatch conditions between networks and datasets, which we refer to as half- and complete-mismatch scenarios. To clarify more, a half mismatch indicates that the SN and TN are trained with the same architectures or datasets. However, complete mismatches indicate SN and TN, and the datasets are completely different. In this study, we considered three different scenarios in transferability, namely: \textit{Cross-Training}, \textit{Cross-Model}, and \textit{Cross Model and Training}.
%
\begin{itemize}
    \item \textit{\textbf{Cross-Training}:} The SN and TN in this half-mismatch scenario have the same architectural frameworks, but they were trained on two different datasets: CIC and UNSW.
    \item \textit{\textbf{Cross-Model}:} This scenario likewise refers to a half-mismatch, which is a mismatch in the networks \textit{QUB1} and \textit{QUB2}, but matches in datasets.
    \item \textit{\textbf{Cross Model and Training}:} In this case, SN and TN are completely mismatched because they have different network architectures and are trained on different datasets.
\end{itemize}

The three scenarios resulted in a significant number of examinations, the results of which are discussed in Section \ref{Results}. To keep our experiments simple and understandable, we did not cover all potential possibilities. However, we determined that an improvement in transferability was exhibited after an appropriate number of analyses.

\subsection{Improving attack strength}
\label{Imp_Att}

Recent studies on multiple domains have demonstrated that most adversarial attacks are naturally non-transferable between the SN and TN \cite{Ding2021ATransferability}. Therefore, in this study, our main goal is to increase the strength of the attack capability that transfers between networks (here from SN to TN), especially in CNN, owing to the popularity of this network and its performance. 
Furthermore, attack transferability must be evaluated in scenarios that differ from the standard implementations of attacks outlined in most adversarial software packages, such as Foolbox library toolkits \cite{FoolBox}.  
%
\begin{table*}[!h]
\scriptsize
\centering
\caption{Cross-training transferability experimental results: mismatch in datasets but matching in \textit{QUB1} networks.\\
\label{CT_QUB1}}
\resizebox{1.8\columnwidth}{!}{
\begin{tabular}{|c|c|c|c|c|c|c|c|}
\hline
\textbf{SN} & \textbf{TN} & \textbf{Attack Type} & \textbf{PSNR} & \textbf{$L_1$ dist} & \textbf{Max. dist} & \textbf{ASR (SN)} & \textbf{ASR (TN)}  \\ \hline

 \rowcolor{grey}
$N_{QUB1}^{UNSW}$ & $N_{QUB1}^{CIC}$ & I-FGSM, $\varepsilon$ = 0.1 & 41.11 & 2.00 & 2.55  & 1.0000 & 0.9000 \TBstrut\\ \hline

$N_{QUB1}^{UNSW}$ & $N_{QUB1}^{CIC}$ & I-FGSM, $\varepsilon$ = 0.01 & 40.27 & 2.04 & 3.56 & 1.0000 & 0.8900 \TBstrut\\ \hline

 \rowcolor{grey}
$N_{QUB1}^{UNSW}$ & $N_{QUB1}^{CIC}$ & I-FGSM, $\varepsilon$ = 0.001 & 40.08 & 2.08 & 3.69 & 1.0000 & 0.8901 \TBstrut\\ \hline

$N_{QUB1}^{UNSW}$ & $N_{QUB1}^{CIC}$ & FGSM, $\varepsilon$ = 0.1 & 41.11 & 2.00 & 2.55 & 1.0000 & 0.9000 \TBstrut\\ \hline

\rowcolor{grey}
$N_{QUB1}^{UNSW}$ & $N_{QUB1}^{CIC}$ & FGSM, $\varepsilon$ = 0.01 & 40.71 & 2.10 & 2.66 & 1.0000 & 0.8200 \TBstrut\\ \hline

$N_{QUB1}^{UNSW}$ & $N_{QUB1}^{CIC}$ & FGSM, $\varepsilon$ = 0.001 & 41.07 & 2.04 & 2.57 & 1.0000 & 0.8100 \TBstrut\\ \hline

\rowcolor{grey}
$N_{QUB1}^{UNSW}$ & $N_{QUB1}^{CIC}$ & JSMA, $\theta$ = 0.1 & 41.06 & 0.07 & 93.47 & 1.0000 & 0.8300 \TBstrut\\ \hline

$N_{QUB1}^{UNSW}$ & $N_{QUB1}^{CIC}$ & JSMA, $\theta$ = 0.01 & 44.07 & 0.16 & 17.85 & 1.0000 & 0.7100 \TBstrut\\ \hline

\rowcolor{grey}
$N_{QUB1}^{UNSW}$ & $N_{QUB1}^{CIC}$ & LBFGS, default parameter & 15.48 & 31.52 & 205.10 & 1.0000 & 0.7100 \TBstrut\\ \hline

$N_{QUB1}^{UNSW}$ & $N_{QUB1}^{CIC}$ & PGD, default parameter & 40.90 & 1.90 & 3.32 & 1.0000 & 0.6300 \TBstrut\\ \hline

\rowcolor{grey}
$N_{QUB1}^{UNSW}$ & $N_{QUB1}^{CIC}$ & CW, $c$ = 100 & 48.95 & 0.24 & 15.32 & 1.0000 & 0.8500 \TBstrut\\ \hline

$N_{QUB1}^{CIC}$ & $N_{QUB1}^{UNSW}$ & I-FGSM, $\varepsilon$ = 0.1 & 41.94 & 1.65 & 2.55  & 1.0000 & 0.9700 \TBstrut\\ \hline

\rowcolor{grey}
$N_{QUB1}^{CIC}$ & $N_{QUB1}^{UNSW}$ & I-FGSM, $\varepsilon$ = 0.01 & 40.51 & 1.88 & 3.59  & 1.0000 & 0.7300 \TBstrut\\ \hline

$N_{QUB1}^{CIC}$ & $N_{QUB1}^{UNSW}$ & I-FGSM, $\varepsilon$ = 0.001 & 40.03 & 1.97 & 3.83  & 1.0000 & 0.7300 \TBstrut\\ \hline

\rowcolor{grey}
$N_{QUB1}^{CIC}$ & $N_{QUB1}^{UNSW}$ & FGSM, $\varepsilon$ = 0.1 & 41.11 & 2.00 & 2.55 & 1.0000 & 0.9000 \TBstrut\\ \hline

$N_{QUB1}^{CIC}$ & $N_{QUB1}^{UNSW}$ & FGSM, $\varepsilon$ = 0.01 & 40.71 & 2.10 & 2.66 & 1.0000 & 0.8200 \TBstrut\\ \hline

\rowcolor{grey}
$N_{QUB1}^{CIC}$ & $N_{QUB1}^{UNSW}$ & FGSM, $\varepsilon$ = 0.001 & 41.07 & 2.04 & 2.57 & 1.0000 & 0.8100 \TBstrut\\ \hline

$N_{QUB1}^{CIC}$ & $N_{QUB1}^{UNSW}$ & JSMA, $\theta$ = 0.1 & 50.88 & 0.03 & 25.49 & 1.0000 & 0.9800 \TBstrut\\ \hline

\rowcolor{grey}
$N_{QUB1}^{CIC}$ & $N_{QUB1}^{UNSW}$ & JSMA, $\theta$ = 0.01 & 53.59 & 0.03 & 13.00 & 1.0000 & 0.9800 \TBstrut\\ \hline

$N_{QUB1}^{CIC}$ & $N_{QUB1}^{UNSW}$ & LBFGS, default parameter & 21.20 & 12.27 & 195.85 & 1.0000 & 0.5200 \TBstrut\\ \hline

\rowcolor{grey}
$N_{QUB1}^{CIC}$ & $N_{QUB1}^{UNSW}$ & PGD, default parameter & 41.49 & 1.69 & 3.12 & 1.0000 & 0.7300 \TBstrut\\ \hline

$N_{QUB1}^{CIC}$ & $N_{QUB1}^{UNSW}$ & CW, $c$ = 100 & 45.56 & 0.31 & 14.32 & 1.0000 & 0.7600 \TBstrut\\ \hline

\end{tabular}}
\end{table*}

\begin{table*}[!h]
\scriptsize
\centering
\caption{Cross-training transferability experimental results: mismatch in datasets but matching in \textit{QUB2} networks. \\
\label{CT_QUB2}}
\resizebox{1.8\columnwidth}{!}{
\begin{tabular}{|c|c|c|c|c|c|c|c|}
\hline
\textbf{SN} & \textbf{TN} & \textbf{Attack Type} & \textbf{PSNR} & \textbf{$L_1$ dist} & \textbf{Max. dist} & \textbf{ASR (SN)} & \textbf{ASR (TN)}  \\ \hline

 \rowcolor{grey}
$N_{QUB2}^{UNSW}$ & $N_{QUB2}^{CIC}$ & I-FGSM, $\varepsilon$ = 0.1 & 35.58 & 2.91 & 10.07  & 1.0000 & 0.9800 \TBstrut\\ \hline

$N_{QUB2}^{UNSW}$ & $N_{QUB2}^{CIC}$ & I-FGSM, $\varepsilon$ = 0.01 & 35.94 & 2.79 & 9.64 & 1.0000 & 0.9800 \TBstrut\\ \hline

 \rowcolor{grey}
$N_{QUB2}^{UNSW}$ & $N_{QUB2}^{CIC}$ & I-FGSM, $\varepsilon$ = 0.001 & 36.00 & 2.77 & 9.53 & 1.0000 & 1.0000 \TBstrut\\ \hline

$N_{QUB2}^{UNSW}$ & $N_{QUB2}^{CIC}$ & FGSM, $\varepsilon$ = 0.1 & 25.79 & 10.12 & 17.74 & 1.0000 & 1.0000 \TBstrut\\ \hline

\rowcolor{grey}
$N_{QUB2}^{UNSW}$ & $N_{QUB2}^{CIC}$ & FGSM, $\varepsilon$ = 0.01 & 25.87 & 10.04 & 17.59 & 1.0000 & 1.0000 \TBstrut\\ \hline

$N_{QUB2}^{UNSW}$ & $N_{QUB2}^{CIC}$ & FGSM, $\varepsilon$ = 0.001 & 25.97 & 9.92 & 17.74 & 1.0000 & 1.0000 \TBstrut\\ \hline

\rowcolor{grey}
$N_{QUB2}^{UNSW}$ & $N_{QUB2}^{CIC}$ & JSMA, $\theta$ = 0.1 & 32.41 & 0.36 & 178.50 & 1.0000 & 0.6700 \TBstrut\\ \hline

$N_{QUB2}^{UNSW}$ & $N_{QUB2}^{CIC}$ & JSMA, $\theta$ = 0.01 & 34.07 & 0.38 & 181.05 & 1.0000 & 0.4300 \TBstrut\\ \hline

\rowcolor{grey}
$N_{QUB2}^{UNSW}$ & $N_{QUB2}^{CIC}$ & LBFGS, default parameter & 24.90 & 7.20 & 135.18 & 1.0000 & 0.8100 \TBstrut\\ \hline

$N_{QUB2}^{UNSW}$ & $N_{QUB2}^{CIC}$ & PGD, default parameter & 35.79 & 2.87 & 9.27 & 1.0000 & 0.9100 \TBstrut\\ \hline

\rowcolor{grey}
$N_{QUB2}^{UNSW}$ & $N_{QUB2}^{CIC}$ & CW, $c$ = 100 & 40.19 & 0.90 & 31.59 & 1.0000 & 0.6300 \TBstrut\\ \hline


$N_{QUB2}^{CIC}$ & $N_{QUB2}^{UNSW}$ & I-FGSM, $\varepsilon$ = 0.1 & 40.38 & 1.81 & 4.32 & 1.0000 & 0.5400 \TBstrut\\ \hline

\rowcolor{grey}
$N_{QUB2}^{CIC}$ & $N_{QUB2}^{UNSW}$ & I-FGSM, $\varepsilon$ = 0.01 & 40.05 & 1.89 & 4.51 & 1.0000 & 0.4700 \TBstrut\\ \hline

$N_{QUB2}^{CIC}$ & $N_{QUB2}^{UNSW}$ & I-FGSM, $\varepsilon$ = 0.001 & 40.02 & 1.90 & 4.52  & 1.0000 & 0.4700 \TBstrut\\ \hline

\rowcolor{grey}
$N_{QUB2}^{UNSW}$ & $N_{QUB2}^{CIC}$ & FGSM, $\varepsilon$ = 0.1 & 44.70 & 1.32 & 1.90 & 1.0000 & 0.8600 \TBstrut\\ \hline

$N_{QUB2}^{UNSW}$ & $N_{QUB2}^{CIC}$ & FGSM, $\varepsilon$ = 0.01 & 41.62 & 1.76 & 2.57 & 1.0000 & 0.8500 \TBstrut\\ \hline

\rowcolor{grey}
$N_{QUB2}^{UNSW}$ & $N_{QUB2}^{CIC}$ & FGSM, $\varepsilon$ = 0.001 & 45.53 & 0.20 & 1.71 & 1.0000 & 0.8500 \TBstrut\\ \hline

$N_{QUB2}^{CIC}$ & $N_{QUB2}^{UNSW}$ & JSMA, $\theta$ = 0.1 & 42.93 & 0.23 & 41.87 & 1.0000 & 0.9900 \TBstrut\\ \hline

\rowcolor{grey}
$N_{QUB2}^{CIC}$ & $N_{QUB2}^{UNSW}$ & JSMA, $\theta$ = 0.01 & 42.86 & 0.27 & 44.55 & 1.0000 & 0.8800 \TBstrut\\ \hline

$N_{QUB2}^{CIC}$ & $N_{QUB2}^{UNSW}$ & LBFGS, default parameter & 23.87 & 7.34 & 140.10 & 1.0000 & 0.9100 \TBstrut\\ \hline

\rowcolor{grey}
$N_{QUB2}^{CIC}$ & $N_{QUB2}^{UNSW}$ & PGD, default parameter & 41.27 & 1.69 & 3.45 & 1.0000 & 0.4400 \TBstrut\\ \hline

$N_{QUB2}^{CIC}$ & $N_{QUB2}^{UNSW}$ & CW, $c$ = 100 & 41.20 & 0.81 & 34.60 & 1.0000 & 0.7100 \TBstrut\\ \hline

\end{tabular}}
\end{table*}

Adversarial frameworks, such as Foolbox, are primarily developed to minimize embedding distortion and facilitate successful adversarial attacks. Therefore, the attack examples in most adversarial toolkits are close to the detection border; thus, even small modifications to the detector settings might compromise the efficacy of the attack. As a result, we developed this method by modifying software packages such as Foolbox to increase the transferability of attacks between networks when an attack example travels more inside the attack's target region but maintains a distortion as low as possible. Although we applied our strategy in this toolbox, it can also be applied to other available toolboxes. An attack procedure is outlined in Algorithm \ref{algo}. In this Algorithm, $x$ and $y$ are related to the original sample and the corresponding sample labels, and $SN(x)$ and $TN(x)$ are related to the source and target networks. As previously mentioned, two hyperparameters play important roles in the algorithm, one of which is related to the maximum distortion ($\epsilon$) that should be applied and the boundary distance ($\delta$). $x'$ is an adversarial attack that is derived from the original sample $x$ and the goal is to minimize $\lVert x - x' \rVert$ while ensuring that the sample is misclassified from $SN$. However, in this case, the sample is close to the decision margin. Then,  $x''$ is an updated version of $x'$ that intends to be transferred between $SN$ and $TN$.

\noindent \textbf{Algorithm: Increasing Attack Transferability} \label{algo}

\begin{enumerate}
    \item \textbf{Input:} ($x$, $y$), $SN(x)$, and $TN(x)$
    \item \textbf{Hyperparameters:} ($\epsilon$, and $\delta$)
    \item \textbf{Output:} 
          \begin{itemize}
              \item Adversarial example $x''$ that transfers from $SN(x)$ to $TN(x)$
          \end{itemize}
    \item Initialize $x' = x$
    \item \textbf{while} $SN(x') = y$ and $TN(x') = y$ \textbf{do}
          \begin{itemize}
              \item Use attack method (FGSM, PGD, etc) to generate $x'$ that minimizes $\lVert x - x' \rVert$
          \end{itemize}
    \item \textbf{while} $\lVert SN(x') - y \rVert < \delta$ \textbf{do}
          \begin{itemize}
              \item Modify $x'$ to move farther from $SN(x)$ decision boundary
              \item Subject to: $\lVert x - x' \rVert < \epsilon$
          \end{itemize}
    \item Return adversarial example $x'' = x'$
\end{enumerate}

\section{Experiential exploration and Discussion}
\label{Results}
In this section, we outline the effectiveness of our approach, which is utilized to enhance the transferability of adversarial examples in various scenarios, and provide our experimental findings.

\subsection{Experiential Exploration}
The test accuracies of \textit{QUB1} and \textit{QUB2} on the CIC and UNSW test sets are as follows: $N_{QUB1}^{CIC}$ demonstrated an accuracy of 97.00\%, $N_{QUB1}^{UNSW}$ achieved 100\% accuracy, $N_{QUB2}^{CIC}$ achieved 97.72\% accuracy, and $N_{QUB2}^{UNSW}$ achieved an ideal accuracy of 100\%.\\
\indent Regarding the I-FGSM adversarial attack, the steps were consistently set to 10, while diverse normalized strength factors were considered, that is, $\varepsilon = 0.1, 0.01, 0.001$. The FGSM attacks employed the same strength factor. For JSMA, the model parameters $\theta$ were set to values of 0.1 and 0.01. The default strength factor parameter was employed for LBFGS and PGD attacks. Finally, a confidence parameter of 100 was used for $c$ in a CW adversarial attack.
The reported outcomes include the mean attack success rate (ASR) on SN and TN, mean peak signal-to-noise ratio (PSNR) estimated within 500 samples, mean $L_1$ distortion, and mean maximum absolute distortion. In the experimental settings, strengthening the transferability of adversarial attacks was deemed successful if the standard ASR threshold for TN surpassed 40\%.

\textbf{Cross-Training transferability.} 
We explain and present an analysis of cross-training transferability in~\ref{CT_QUB1} and \ref{CT_QUB2}, in which \textit{QUB1} and \textit{QUB2} networks were considered for both SN and TN that trained with CIC and UNSW datasets. 
The average ASR on an SN is typically successful in transferring attacks to the TN. To provide an example, Table~\ref{CT_QUB1} where \textit{QUB1} is considered for SN and TN, for the attacks FGSM and I-FGSM with attack parameter $\varepsilon = 0.1$ the ASR on TN is 0.9000 or 90\%, which means that attacks on SN are sufficient to fool a TN, as well as for other parameters. In another example, JSMA with attack parameters  $\theta = 0.1, 0.01$, the LBFGS, the PGD, and even CW with a parameter $c = 100$, the ASR proves that our methodology can increase the transferability of attacks between SN and TN. Therefore, the transfer of attacks from SN to TN was significant for six adversarial attacks in the \textit{QUB1} network when SN was trained with the UNSW dataset and TN was trained with the CIC dataset. In addition, we see that if SN is trained with CIC and TN is trained with the UNSW dataset, the ASR on TN proves that attacks on SN are sufficient to transfer to TN by considering our strategy. 
Regarding Table\ref{CT_QUB2}, when the \textit{QUB2} network is considered for SN and TN and both of them are already trained with CIC and UNSW, the ASR on TN is almost higher than 40\% of our threshold, which means that attacks on SN are enough to fool TN.  
In general, improving the strength of the attack strategy improves the transferability of adversarial attacks, enabling the successful deception of the TN. When SN and TN shared the same network but were trained on different datasets, cross-training transferability was found to be asymmetric. Earlier studies demonstrated that most adversarial attacks were not transferable across SN and TN in CNN, but only a few scenarios were, such as the cross-training scenario \cite{Nowroozi2022Dem}. 
Nonetheless, we confirmed that improving the attack strength can improve the transferability of adversarial attacks between SN and TN. This can be attributed to attacks on a larger percentage of the examples.

\textbf{Cross-Model transferability.} We study manifold scenarios regarding cross-model transferability and demonstrate the corresponding experiments in Tables ~\ref{CMT_First} and ~\ref{CMT_Second} while employing two specific networks (\textit{QUB1} and \textit{QUB2}) when they are trained with the same datasets.
\begin{table*}[!t]
\scriptsize
\centering
\caption{Cross-model transferability experimental results: mismatch in networks but matching in UNSW datasets.\\ \label{CMT_First}}
\resizebox{1.8\columnwidth}{!}{
\begin{tabular}{|c|c|c|c|c|c|c|c|}
\hline
\textbf{SN} & \textbf{TN} & \textbf{Attack Type} & \textbf{PSNR} & \textbf{$L_1$ dist} & \textbf{Max. dist} & \textbf{ASR (SN)} & \textbf{ASR (TN)} \\ \hline

 \rowcolor{grey}
$N_{QUB1}^{UNSW}$ & $N_{QUB2}^{UNSW}$ & I-FGSM, $\varepsilon$ = 0.1 & 40.43 & 2.02 & 3.44  & 1.0000 & 0.8000 \TBstrut\\ \hline

$N_{QUB1}^{UNSW}$ & $N_{QUB2}^{UNSW}$ & I-FGSM, $\varepsilon$ = 0.01 & 40.07 & 2.09 & 3.66  & 1.0000 & 0.7400 \TBstrut\\ \hline

 \rowcolor{grey}
$N_{QUB1}^{UNSW}$ & $N_{QUB2}^{UNSW}$ & I-FGSM, $\varepsilon$ = 0.001 & 40.04 & 2.10 & 3.68  & 1.0000 & 0.4100 \TBstrut\\ \hline

$N_{QUB1}^{UNSW}$ & $N_{QUB2}^{UNSW}$ & FGSM, $\varepsilon$ = 0.1 & 40.98 & 2.05 & 2.55  & 1.0000 & 0.7700 \TBstrut\\ \hline

\rowcolor{grey}
$N_{QUB1}^{UNSW}$ & $N_{QUB2}^{UNSW}$ & FGSM, $\varepsilon$ = 0.01 & 40.69 & 2.12 & 2.64  & 1.0000 & 0.6300 \TBstrut\\ \hline

$N_{QUB1}^{UNSW}$ & $N_{QUB2}^{UNSW}$ & FGSM, $\varepsilon$ = 0.001 & 41.30 & 1.99 & 2.48 & 1.0000 & 0.6300 \TBstrut\\ \hline

\rowcolor{grey}
$N_{QUB1}^{UNSW}$ & $N_{QUB2}^{UNSW}$ & JSMA, $\theta$ = 0.1 & 41.39 & 0.06 & 90.38  & 1.0000 & 0.6100 \TBstrut\\ \hline

$N_{QUB1}^{UNSW}$ & $N_{QUB2}^{UNSW}$ & JSMA, $\theta$ = 0.01 & 41.81 & 0.06 & 90.03 & 1.0000 & 0.5900 \TBstrut\\ \hline

\rowcolor{grey}
$N_{QUB1}^{UNSW}$ & $N_{QUB2}^{UNSW}$ & LBFGS, default parameter & 25.10 & 24.06 & 110.30 & 1.0000 & 0.7100 \TBstrut\\ \hline

$N_{QUB1}^{UNSW}$ & $N_{QUB2}^{UNSW}$ & PGD, default parameter & 41.20 & 1.85 & 3.13 & 1.0000 & 0.6400 \TBstrut\\ \hline

\rowcolor{grey}
$N_{QUB1}^{UNSW}$ & $N_{QUB2}^{UNSW}$ & CW, $c$ = 100 & 40.19 & 0.85 & 28.72 & 1.0000 & 0.8000 \TBstrut\\ \hline

%

$N_{QUB2}^{UNSW}$ & $N_{QUB1}^{UNSW}$ & I-FGSM, $\varepsilon$ = 0.1 & 35.97 & 2.76 & 9.63  & 1.0000 & 0.8800 \TBstrut\\ \hline

\rowcolor{grey}
$N_{QUB2}^{UNSW}$ & $N_{QUB1}^{UNSW}$ & I-FGSM, $\varepsilon$ = 0.01 & 36.20 & 2.68 & 9.33  & 1.0000 & 0.8200 \TBstrut\\ \hline

$N_{QUB2}^{UNSW}$ & $N_{QUB1}^{UNSW}$ & I-FGSM, $\varepsilon$ = 0.001 & 36.18 & 2.69 & 9.37 & 1.0000 & 0.8300 \TBstrut\\ \hline

 \rowcolor{grey}
$N_{QUB2}^{UNSW}$ & $N_{QUB1}^{UNSW}$ & FGSM, $\varepsilon$ = 0.1 & 25.58 & 10.30 & 18.44  & 1.0000 & 1.0000 \TBstrut\\ \hline

$N_{QUB2}^{UNSW}$ & $N_{QUB1}^{UNSW}$ & FGSM, $\varepsilon$ = 0.01 & 25.67 & 10.20 & 18.24  & 1.0000 & 1.0000 \TBstrut\\ \hline

 \rowcolor{grey}
$N_{QUB2}^{UNSW}$ & $N_{QUB1}^{UNSW}$ & FGSM, $\varepsilon$ = 0.001 & 25.79 & 10.06 & 17.98 & 1.0000 & 1.0000 \TBstrut\\ \hline

$N_{QUB2}^{UNSW}$ & $N_{QUB1}^{UNSW}$ & JSMA, $\theta$ = 0.1 & 32.17 & 0.32 & 226.76 & 1.0000 & 0.8700 \TBstrut\\ \hline

 \rowcolor{grey}
$N_{QUB2}^{UNSW}$ & $N_{QUB1}^{UNSW}$ & JSMA, $\theta$ = 0.01 & 32.60 & 0.31 & 222.93 & 1.0000 & 0.8500 \TBstrut\\ \hline

$N_{QUB2}^{UNSW}$ & $N_{QUB1}^{UNSW}$ & LBFGS, default parameter & 24.10 & 22.27 & 124.33 & 1.0000 & 0.8700 \TBstrut\\ \hline

 \rowcolor{grey}
$N_{QUB2}^{UNSW}$ & $N_{QUB1}^{UNSW}$ & PGD, default parameter & 36.00 & 2.77 & 9.15 & 1.0000 & 0.8100 \TBstrut\\ \hline

$N_{QUB2}^{UNSW}$ & $N_{QUB1}^{UNSW}$ & CW, $c$ = 100 & 40.38 & 0.80 & 29.76 & 1.0000 & 0.8000 \TBstrut\\ \hline

\end{tabular}}
\end{table*}

\begin{table*}[!h]
\scriptsize
\centering
\caption{Cross-model transferability experimental results: mismatch in networks but matching in CIC datasets.\\. \label{CMT_Second}}
\resizebox{1.8\columnwidth}{!}{
\begin{tabular}{|c|c|c|c|c|c|c|c|}
\hline
\textbf{SN} & \textbf{TN} & \textbf{Attack Type} & \textbf{PSNR} & \textbf{$L_1$ dist} & \textbf{Max. dist} & \textbf{ASR (SN)} & \textbf{ASR (TN)}  \\ \hline

 \rowcolor{grey}
$N_{QUB1}^{CIC}$ & $N_{QUB2}^{CIC}$ & I-FGSM, $\varepsilon$ = 0.1 & 42.02 & 1.64 & 2.57  & 1.0000 & 0.8300 \TBstrut\\ \hline

$N_{QUB1}^{CIC}$ & $N_{QUB2}^{CIC}$ & I-FGSM, $\varepsilon$ = 0.01 & 40.14 & 2.08 & 3.46  & 1.0000 & 0.6900 \TBstrut\\ \hline

 \rowcolor{grey}
$N_{QUB1}^{CIC}$ & $N_{QUB2}^{CIC}$ & I-FGSM, $\varepsilon$ = 0.001 & 40.03 & 2.11 & 3.51 & 1.0000 & 0.6700 \TBstrut\\ \hline

$N_{QUB1}^{CIC}$ & $N_{QUB2}^{CIC}$ & FGSM, $\varepsilon$ = 0.1 & 41.87 & 1.85 & 2.40 & 1.0000 & 0.8700 \TBstrut\\ \hline

\rowcolor{grey}
$N_{QUB1}^{CIC}$ & $N_{QUB2}^{CIC}$ & FGSM, $\varepsilon$ = 0.01 & 41.01 & 2.00 & 2.61 & 1.0000 & 0.8200 \TBstrut\\ \hline

$N_{QUB1}^{CIC}$ & $N_{QUB2}^{CIC}$ & FGSM, $\varepsilon$ = 0.001 & 43.42 & 1.62 & 2.11 & 1.0000 & 0.7800 \TBstrut\\ \hline

\rowcolor{grey}
$N_{QUB1}^{CIC}$ & $N_{QUB2}^{CIC}$ & JSMA, $\theta$ = 0.1 & 44.72 & 0.20 & 39.61 & 1.0000 & 0.9200 \TBstrut\\ \hline

$N_{QUB1}^{CIC}$ & $N_{QUB2}^{CIC}$ & JSMA, $\theta$ = 0.01 & 44.21 & 0.22 & 42.37 & 1.0000 & 0.9000 \TBstrut\\ \hline

\rowcolor{grey}
$N_{QUB1}^{CIC}$ & $N_{QUB2}^{CIC}$ & LBFGS, default parameter & 32.01 & 4.42 & 67.67 & 1.0000 & 0.4200 \TBstrut\\ \hline

$N_{QUB1}^{CIC}$ & $N_{QUB2}^{CIC}$ & PGD, default parameter & 40.11 & 1.80 & 2.13 & 1.0000 & 0.5000 \TBstrut\\ \hline

\rowcolor{grey}
$N_{QUB1}^{CIC}$ & $N_{QUB2}^{CIC}$ & CW, $c$ = 100 & 41.89 & 0.91 & 16.33 & 1.0000 & 0.4400 \TBstrut\\ \hline

%

$N_{QUB2}^{CIC}$ & $N_{QUB1}^{CIC}$ & I-FGSM, $\varepsilon$ = 0.1 & 40.63 & 1.80 & 3.82  & 1.0000 & 0.6300 \TBstrut\\ \hline

\rowcolor{grey}
$N_{QUB2}^{CIC}$ & $N_{QUB1}^{CIC}$ & I-FGSM, $\varepsilon$ = 0.01 & 40.17 & 1.87 & 4.44  & 1.0000 & 0.5200 \TBstrut\\ \hline

$N_{QUB2}^{CIC}$ & $N_{QUB1}^{CIC}$ & I-FGSM, $\varepsilon$ = 0.001 & 40.07 & 1.89 & 4.50 & 1.0000 & 0.5100 \TBstrut\\ \hline

 \rowcolor{grey}
$N_{QUB2}^{CIC}$ & $N_{QUB1}^{CIC}$ & FGSM, $\varepsilon$ = 0.1 & 44.23 & 1.39 & 1.99 & 1.0000 & 0.8600 \TBstrut\\ \hline

$N_{QUB2}^{CIC}$ & $N_{QUB1}^{CIC}$ & FGSM, $\varepsilon$ = 0.01 & 41.62 & 1.76 & 2.56 & 1.0000 & 0.9100 \TBstrut\\ \hline

 \rowcolor{grey}
$N_{QUB2}^{CIC}$ & $N_{QUB1}^{CIC}$ & FGSM, $\varepsilon$ = 0.001 & 45.22 & 1.25 & 1.78 & 1.0000 & 0.8400 \TBstrut\\ \hline

$N_{QUB2}^{CIC}$ & $N_{QUB1}^{CIC}$ & JSMA, $\theta$ = 0.1 & 42.32 & 0.24 & 41.13 & 1.0000 & 0.9800 \TBstrut\\ \hline

 \rowcolor{grey}
$N_{QUB2}^{CIC}$ & $N_{QUB1}^{CIC}$ & JSMA, $\theta$ = 0.01 & 42.90 & 0.28 & 41.83 & 1.0000 & 0.8600 \TBstrut\\ \hline

$N_{QUB2}^{CIC}$ & $N_{QUB1}^{CIC}$ & LBFGS, default parameter & 25.94 & 13.39 & 72.99 & 1.0000 & 0.9700 \TBstrut\\ \hline

 \rowcolor{grey}
$N_{QUB2}^{CIC}$ & $N_{QUB1}^{CIC}$ & PGD, default parameter & 41.04 & 1.76 & 3.42 & 1.0000 & 0.5200 \TBstrut\\ \hline

$N_{QUB2}^{CIC}$ & $N_{QUB1}^{CIC}$ & CW, $c$ = 100 & 45.55 & 0.81 & 12.29 & 1.0000 & 0.5100 \TBstrut\\ \hline

\end{tabular}}
\end{table*}

%
We note that our analyses show that the average ASR on the SN is 1.0000 or 100\%; however, the average ASR on the TN exceeds the 40\% threshold, showing that attacks travel entirely from the SN to the TN after applying our strategy.
Concerning the UNSW, the I-FGSM attack exhibited a robust transferability of 80\% from the SN to the TN when assessing the SN with $N_{QUB1}^{UNSW}$ and $N_{QUB2}^{UNSW}$ using a parameter value of $\varepsilon = 0.1$. 
%
Additionally, when exploring an SN with $N_{QUB2}^{UNSW}$ and $N_{QUB1}^{UNSW}$ using the same parameters, it demonstrates a significant transferability of 88\%.
When analyzing the SN by employing $N_{QUB1}^{CIC}$ and $N_{QUB2}^{CIC}$ with a parameter value of $theta = 0.1$, the JSMA attack exhibited a significant transferability of 92\%.
Correspondingly, it exhibits an outstanding transferability of 98\% from the SN to the TN when evaluating an SN with $N_{QUB2}^{CIC}$ and $N_{QUB1}^{CIC}$ using the same parameter.
In contrast, in the aforementioned cross-training scenario, cross-model transferability reveals distinct possibilities, where the attacker can successfully deceive the TN. This outcome was expected because the datasets in the cross-model were identical. Hence, the transferability of an attack in a cross-modal transferability scenario plays a significant role in adversarial attack transferability. Moreover, when shallow and deep architectures are trained with different datasets, peculiar features emerge, potentially indicating the failure of the proposed method. This statement was evident in the cross-training scenario.

\textbf{Cross-Model and -Training transferability.} 
\begin{table*}[!h]
\scriptsize
\centering
\caption{Cross-model and -training transferability experimental results:  QUB1 is considered for SN and QUB2  for TN and trained with different datasets UNSW and CIC.\\ \label{CMTT}}
\resizebox{1.8\columnwidth}{!}{
\begin{tabular}{|c|c|c|c|c|c|c|c|}
\hline
\textbf{SN} & \textbf{TN} & \textbf{Attack Type} & \textbf{PSNR} & \textbf{$L_1$ dist} & \textbf{Max. dist} & \textbf{ASR (SN)} & \textbf{ASR (TN)}   \\ \hline

 \rowcolor{grey}
$N_{QUB1}^{UNSW}$ & $N_{QUB2}^{CIC}$ & I-FGSM, $\varepsilon$ = 0.1 & 39.32 & 2.29 & 3.90  & 1.0000 & 0.7300 \TBstrut\\ \hline

$N_{QUB1}^{UNSW}$ & $N_{QUB2}^{CIC}$ & I-FGSM, $\varepsilon$ = 0.01 & 40.92 & 1.90 & 3.31  & 1.0000 & 0.6700 \TBstrut\\ \hline

 \rowcolor{grey}
$N_{QUB1}^{UNSW}$ & $N_{QUB2}^{CIC}$ & I-FGSM, $\varepsilon$ = 0.001 & 41.14 & 1.86 & 3.24 & 1.0000 & 0.7000 \TBstrut\\ \hline

$N_{QUB1}^{UNSW}$ & $N_{QUB2}^{CIC}$ & FGSM, $\varepsilon$ = 0.1 & 41.14 & 1.99 & 2.55  & 1.0000 & 1.0000 \TBstrut\\ \hline

\rowcolor{grey}
$N_{QUB1}^{UNSW}$ & $N_{QUB2}^{CIC}$ & FGSM, $\varepsilon$ = 0.01 & 40.67 & 2.11 & 2.67  & 1.0000 & 0.9800 \TBstrut\\ \hline

$N_{QUB1}^{UNSW}$ & $N_{QUB2}^{CIC}$ & FGSM, $\varepsilon$ = 0.001 & 39.07 & 2.15 & 2.81 & 1.0000 & 0.9100 \TBstrut\\ \hline

\rowcolor{grey}
$N_{QUB1}^{UNSW}$ & $N_{QUB2}^{CIC}$ & JSMA, $\theta$ = 0.1 & 41.33 & 0.06 & 87.30  & 1.0000 & 0.9100 \TBstrut\\ \hline

$N_{QUB1}^{UNSW}$ & $N_{QUB2}^{CIC}$ & JSMA, $\theta$ = 0.01 & 41.23 & 0.07 & 90.84 & 1.0000 & 0.9000 \TBstrut\\ \hline

\rowcolor{grey}
$N_{QUB1}^{UNSW}$ & $N_{QUB2}^{CIC}$ & LBFGS, default parameter & 41.02 & 1.07 & 23.52 & 1.0000 & 0.5300 \TBstrut\\ \hline

$N_{QUB1}^{UNSW}$ & $N_{QUB2}^{CIC}$ & PGD, default parameter & 40.62 & 1.97 & 3.40 & 1.0000 & 0.6700 \TBstrut\\ \hline

\rowcolor{grey}
$N_{QUB1}^{UNSW}$ & $N_{QUB2}^{CIC}$ & CW, $c$ = 100 & 47.91 & 0.27 & 14.99 & 1.0000 & 0.8000 \TBstrut\\ \hline

%

$N_{QUB1}^{CIC}$ & $N_{QUB2}^{UNSW}$ & I-FGSM, $\varepsilon$ = 0.1 & 42.10 & 1.61 & 2.57 & 1.0000 & 0.7900 \TBstrut\\ \hline

\rowcolor{grey}
$N_{QUB1}^{CIC}$ & $N_{QUB2}^{UNSW}$ & I-FGSM, $\varepsilon$ = 0.01 & 40.14 & 2.07 & 3.52  & 1.0000 & 0.6700 \TBstrut\\ \hline

$N_{QUB1}^{CIC}$ & $N_{QUB2}^{UNSW}$ & I-FGSM, $\varepsilon$ = 0.001 & 40.03 & 2.10 & 3.55 & 1.0000 & 0.6400 \TBstrut\\ \hline

 \rowcolor{grey}
$N_{QUB1}^{CIC}$ & $N_{QUB2}^{UNSW}$ & FGSM, $\varepsilon$ = 0.1 & 41.88 & 1.84 & 2.41  & 1.0000 & 0.8300 \TBstrut\\ \hline

$N_{QUB1}^{CIC}$ & $N_{QUB2}^{UNSW}$ & FGSM, $\varepsilon$ = 0.01 & 41.12 & 1.96 & 2.60  & 1.0000 & 0.7900 \TBstrut\\ \hline

 \rowcolor{grey}
$N_{QUB1}^{CIC}$ & $N_{QUB2}^{UNSW}$ & FGSM, $\varepsilon$ = 0.001 & 43.56 & 1.56 & 2.07 & 1.0000 & 0.7800 \TBstrut\\ \hline

$N_{QUB1}^{CIC}$ & $N_{QUB2}^{UNSW}$ & JSMA, $\theta$ = 0.1 & 46.59 & 0.14 & 33.73 & 1.0000 & 0.9700 \TBstrut\\ \hline

 \rowcolor{grey}
$N_{QUB1}^{CIC}$ & $N_{QUB2}^{UNSW}$ & JSMA, $\theta$ = 0.01 & 47.08 & 0.14 & 32.12 & 1.0000 & 0.9600 \TBstrut\\ \hline

$N_{QUB1}^{CIC}$ & $N_{QUB2}^{UNSW}$ & LBFGS, default parameter & 22.73 & 11.29 & 161.56 & 1.0000 & 0.6200 \TBstrut\\ \hline

 \rowcolor{grey}
$N_{QUB1}^{CIC}$ & $N_{QUB2}^{UNSW}$ & PGD, default parameter & 38.75 & 2.61 & 3.82 & 1.0000 & 0.5400 \TBstrut\\ \hline

$N_{QUB1}^{CIC}$ & $N_{QUB2}^{UNSW}$ & CW, $c$ = 100 & 43.15 & 0.78 & 17.93 & 1.0000 & 0.7300 \TBstrut\\ \hline

\end{tabular}}
\end{table*}

\begin{table*}[!t]
\scriptsize
\centering
\caption{Cross-model and training transferability experimental results:  QUB2 is considered for SN and QUB1  for TN and trained with different datasets UNSW and CIC.\\ \label{CMTTT}}
\resizebox{1.8\columnwidth}{!}{
\begin{tabular}{|c|c|c|c|c|c|c|c|}
\hline
\textbf{SN} & \textbf{TN} & \textbf{Attack Type} & \textbf{PSNR} & \textbf{$L_1$ dist} & \textbf{Max. dist} & \textbf{ASR (SN)} & \textbf{ASR (TN)}  \\ \hline

 \rowcolor{grey}
$N_{QUB2}^{UNSW}$ & $N_{QUB1}^{CIC}$ & I-FGSM, $\varepsilon$ = 0.1 & 35.68 & 2.84 & 10.14  & 1.0000 & 0.7300 \TBstrut\\ \hline

$N_{QUB2}^{UNSW}$ & $N_{QUB1}^{CIC}$ & I-FGSM, $\varepsilon$ = 0.01 & 36.18 & 2.66 & 9.62 & 1.0000 & 0.8200 \TBstrut\\ \hline

 \rowcolor{grey}
$N_{QUB2}^{UNSW}$ & $N_{QUB1}^{CIC}$ & I-FGSM, $\varepsilon$ = 0.001 & 36.01 & 2.73 & 9.77 & 1.0000 & 0.0200 \TBstrut\\ \hline

 \rowcolor{grey}
$N_{QUB2}^{UNSW}$ & $N_{QUB1}^{CIC}$ & FGSM, $\varepsilon$ = 0.1 & 25.57 & 10.19 & 18.78 & 1.0000 & 0.7100 \TBstrut\\ \hline

$N_{QUB2}^{UNSW}$ & $N_{QUB1}^{CIC}$ & FGSM, $\varepsilon$ = 0.01 & 25.65 & 10.10 & 18.60 & 1.0000 & 0.8000 \TBstrut\\ \hline

 \rowcolor{grey}
$N_{QUB2}^{UNSW}$ & $N_{QUB1}^{CIC}$ & FGSM, $\varepsilon$ = 0.001 & 25.77 & 9.96 & 18.32 & 1.0000 & 0.8000 \TBstrut\\ \hline

$N_{QUB2}^{UNSW}$ & $N_{QUB1}^{CIC}$ & JSMA, $\theta$ = 0.1 & 32.24 & 0.32 & 227.12  & 1.0000 & 0.8700 \TBstrut\\ \hline

 \rowcolor{grey}
$N_{QUB2}^{UNSW}$ & $N_{QUB1}^{CIC}$ & JSMA, $\theta$ = 0.01 & 32.66 & 0.31 & 223.29 & 1.0000 & 0.6700 \TBstrut\\ \hline

$N_{QUB2}^{UNSW}$ & $N_{QUB1}^{CIC}$ & LBFGS, default parameter & 32.24 & 0.32 & 227.12 & 1.0000 & 0.8700 \TBstrut\\ \hline

 \rowcolor{grey}
$N_{QUB2}^{UNSW}$ & $N_{QUB1}^{CIC}$ & PGD, default parameter & 36.07 & 2.72 & 9.29 & 1.0000 & 0.4300 \TBstrut\\ \hline

$N_{QUB2}^{UNSW}$ & $N_{QUB1}^{CIC}$ & CW, $c$ = 100 & 40.37 & 0.78 & 29.71 & 1.0000 & 0.7000 \TBstrut\\ \hline

%

\rowcolor{grey}
$N_{QUB2}^{CIC}$ & $N_{QUB1}^{UNSW}$ & I-FGSM, $\varepsilon$ = 0.1 & 41.14 & 1.61 & 3.87 & 1.0000 & 0.9400 \TBstrut\\ \hline

$N_{QUB2}^{CIC}$ & $N_{QUB1}^{UNSW}$ & I-FGSM, $\varepsilon$ = 0.01 & 40.13 & 1.76 & 5.06 & 1.0000 & 0.6700 \TBstrut\\ \hline

 \rowcolor{grey}
$N_{QUB2}^{CIC}$ & $N_{QUB1}^{UNSW}$ & I-FGSM, $\varepsilon$ = 0.001 & 40.11 & 1.77 & 5.03 & 1.0000 & 0.6800 \TBstrut\\ \hline

 \rowcolor{grey}
$N_{QUB2}^{CIC}$ & $N_{QUB1}^{UNSW}$ & FGSM, $\varepsilon$ = 0.1 & 48.04 & 0.80 & 1.27  & 1.0000 & 0.9400 \TBstrut\\ \hline

$N_{QUB2}^{CIC}$ & $N_{QUB1}^{UNSW}$ & FGSM, $\varepsilon$ = 0.01 & 41.99 & 1.60 & 2.60  & 1.0000 & 0.9400 \TBstrut\\ \hline

 \rowcolor{grey}
$N_{QUB2}^{CIC}$ & $N_{QUB1}^{UNSW}$ & FGSM, $\varepsilon$ = 0.001 & 47.08 & 0.96 & 1.53 & 1.0000 & 0.9400 \TBstrut\\ \hline

$N_{QUB2}^{CIC}$ & $N_{QUB1}^{UNSW}$ & JSMA, $\theta$ = 0.1 & 40.57 & 0.41 & 50.99 & 1.0000 & 0.9800 \TBstrut\\ \hline

 \rowcolor{grey}
$N_{QUB2}^{CIC}$ & $N_{QUB1}^{UNSW}$ & JSMA, $\theta$ = 0.01 & 43.40 & 0.28 & 36.33 & 1.0000 & 0.9400 \TBstrut\\ \hline

$N_{QUB2}^{CIC}$ & $N_{QUB1}^{UNSW}$ & LBFGS, default parameter & 22.73 & 11.29 & 161.56 & 1.0000 & 0.6200 \TBstrut\\ \hline

 \rowcolor{grey}
$N_{QUB2}^{CIC}$ & $N_{QUB1}^{UNSW}$ & PGD, default parameter & 17.38 & 26.27 & 146.52 & 1.0000 & 0.7700 \TBstrut\\ \hline

$N_{QUB2}^{CIC}$ & $N_{QUB1}^{UNSW}$ & CW, $c$ = 100 & 41.54 & 1.17 & 19.00 & 1.0000 & 0.6300 \TBstrut\\ \hline

\end{tabular}}
\end{table*}

We evaluate various scenarios in the cross-model and training transferability and present our results in Tables ~\ref{CMTT} ~\ref{CMTTT} which show an assessment of transferability issues of two networks, \textit{QUB1} and \textit{QUB2} when trained on the UNSW and CIC datasets separately. 
This scenario is also one of the complex scenarios for the attacker side because naturally the TN is completely mismatched in the network and dataset compared with SN. Our study shows that the SN has a high average ASR and that many attacks can be successfully transmitted from the SN to the TN, as evidenced by a greater ASR when compared to \cite{Nowroozi2022Dem}'s results, proving that our methodology for increasing the strength of attacks also works in a complex task. Therefore, this transfer is possible in most cases using our methodology. Our findings indicate that it is possible to implement a cross-model and training using our approach, even in cases where there is no cohesion between the datasets and architectures. Consequently, we believe that the transferability of SN and TN can be affirmed confidently. Essentially, our methodology supports this conclusion.

\subsection{Discussion}

Our study led us to conclude that when we enhance the strength of attack capabilities, a large number of adversarial attacks can be transferred between networks.
We achieved this transferability by implementing a new methodology that permits higher distortion in input samples, as long as it results in attacked samples that are located deeper within the target attack region. 
%
Therefore, instances of crafted malicious samples can effectively deceive a different network model even if they have a different architecture.
A crucial observation is that the SN and TN models share similar vulnerabilities, proving their portability. An aggressor seeking to create an effective attack on a TN can use proxy models with various architectures and algorithms. However, there are instances where attacks on TNs do not succeed, despite our method being implemented when considering their ability to adapt to other training. This could be because the decision area is complex and successful attack penetration may cause some regions to lose their significance, whereas others may regain their importance. This component requires deeper examination and signifies a new direction for AI-cybersecurity.
Our findings depict portability, with attention drawn to certain scenarios that showcase its manifestation. In addition, we carried out a series of 500 manipulated examples to approximate the computational expense of several adversary assaults launched on the SN.

\section{Conclusions and Future works}
\label{Conclusions}
By exploring the transferability issue of adversarial CNN-based examples, our study provides an innovative strategy to improve the attack transferability capability of DNNs.
Our approach centers on instances in which the attacker purposely inflates the distortion levels to engender attacked samples that traverse the attack's target region with greater depth. 
%
In this study, we examined different transferability scenarios in the match and mismatch cases between networks and CIC and UNSW datasets.

Observing the cross-training scenario, we noted that only a few attacks did not transfer, because of the complex decision margin. The distortion is challenging to control, which makes it difficult to determine the extent to which the target zone is infiltrated. Sometimes, attempting to re-enter the target model only results in attacking the same area and ending in a total failure. In addition, the level of transferability between CNN-based models and the ideal amount of distortion to consider are influenced by essential factors that are difficult to understand because of the limited transparency of deep neural networks. Transferability is a major concern, particularly for security-oriented uses. Additional investigation is necessary to gain a thorough understanding of the primary factors involved. This knowledge gap must be addressed to ensure success.
Our experiments revealed that, in the cross-training scenario, most attacks exhibited higher transferability than the PGD and CW methods. However, I-FGSM attacks experienced only a few instances of failure in cross-training. These findings are intriguing, as in other scenarios where the majority of attacks are completely transferred between networks, validating the effectiveness of our proposed method. In addition, it is important to note that transferability is not symmetric when considering the training datasets for a specific task. These results suggest that the underlying dataset influences the learned features of the network to some extent even when the proposed method is considered. Consequently, further investigation is required to understand the vulnerability of DL-based network attack transferability.
Moving forward, our future research will focus on fortifying the security of the target model against attacks and exploring ways to control the level of attack penetration and sample distortion inflicted by attackers. However, this remains a complex endeavor owing to the intricate decision boundaries learned by CNNs and other deep neural networks.

%

\bibliographystyle{IEEEtran}
\bibliography{References}

\vskip -2\baselineskip plus -1fil
\begin{IEEEbiography}
[{\includegraphics[width=1in,height=1.25in]{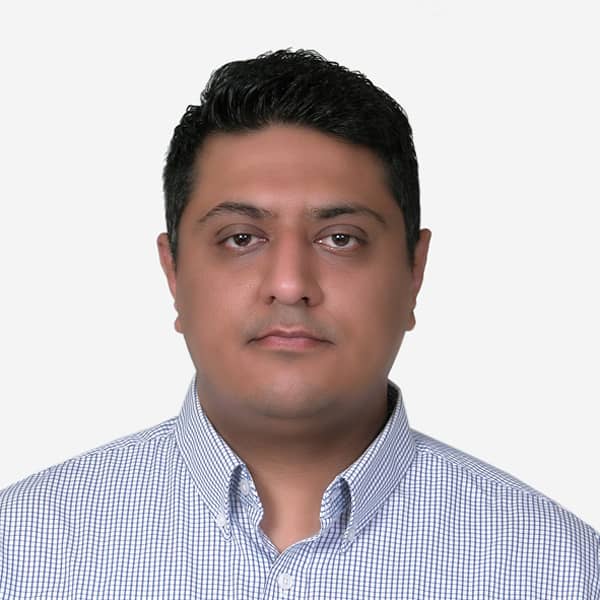}}]
{Ehsan Nowroozi} is a Senior IEEE/ACM Professional Member with expertise in AI for Cybersecurity. He currently serves as a Research Fellow at Queen's University Belfast's Centre for Secure Information Technologies. After earning his Ph.D. in Cybersecurity from the University of Siena in Italy. He holds several postdocs from Siena, Padua, and Sabanci University. He also served as an Assistant Professor at Bahçeşehir University, Turkey, 2022-2023.
\end{IEEEbiography}

\vskip -2\baselineskip plus -1fil
\begin{IEEEbiography}[{\includegraphics[width=1in,height=1.25in]{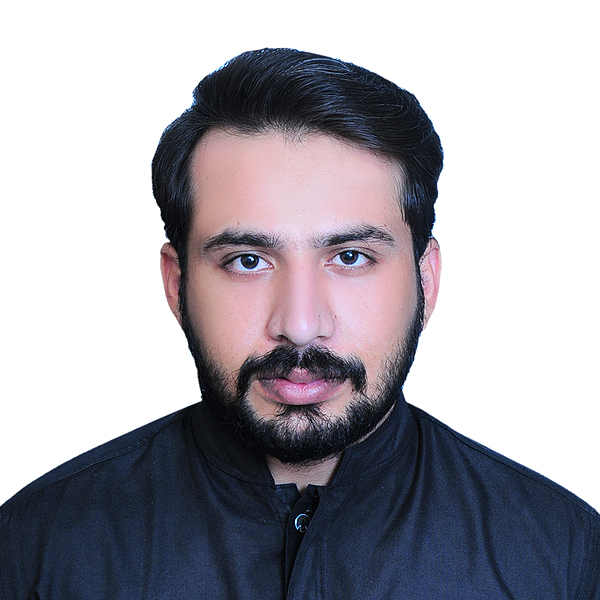}}]{Imran Haider} Imran is a cybersecurity master's student at Bahcesehir University in Istanbul, Turkey. He completed his bachelor’s degree in computer science from the National University of Computer and Emerging Sciences (NUCES), Pakistan. Imran is a student member of IEEE and possesses a strong passion for Cybersecurity and Artificial Intelligence research. In addition to his two years of experience in software development and proficiency in Python for machine/deep learning. he also holds expertise in penetration testing—an essential skill for identifying and mitigating security vulnerabilities in digital systems.
  
\end{IEEEbiography}

\vskip -2\baselineskip plus -1fil
\begin{IEEEbiography}
[{\includegraphics[width=0.9in,height=1.15in]{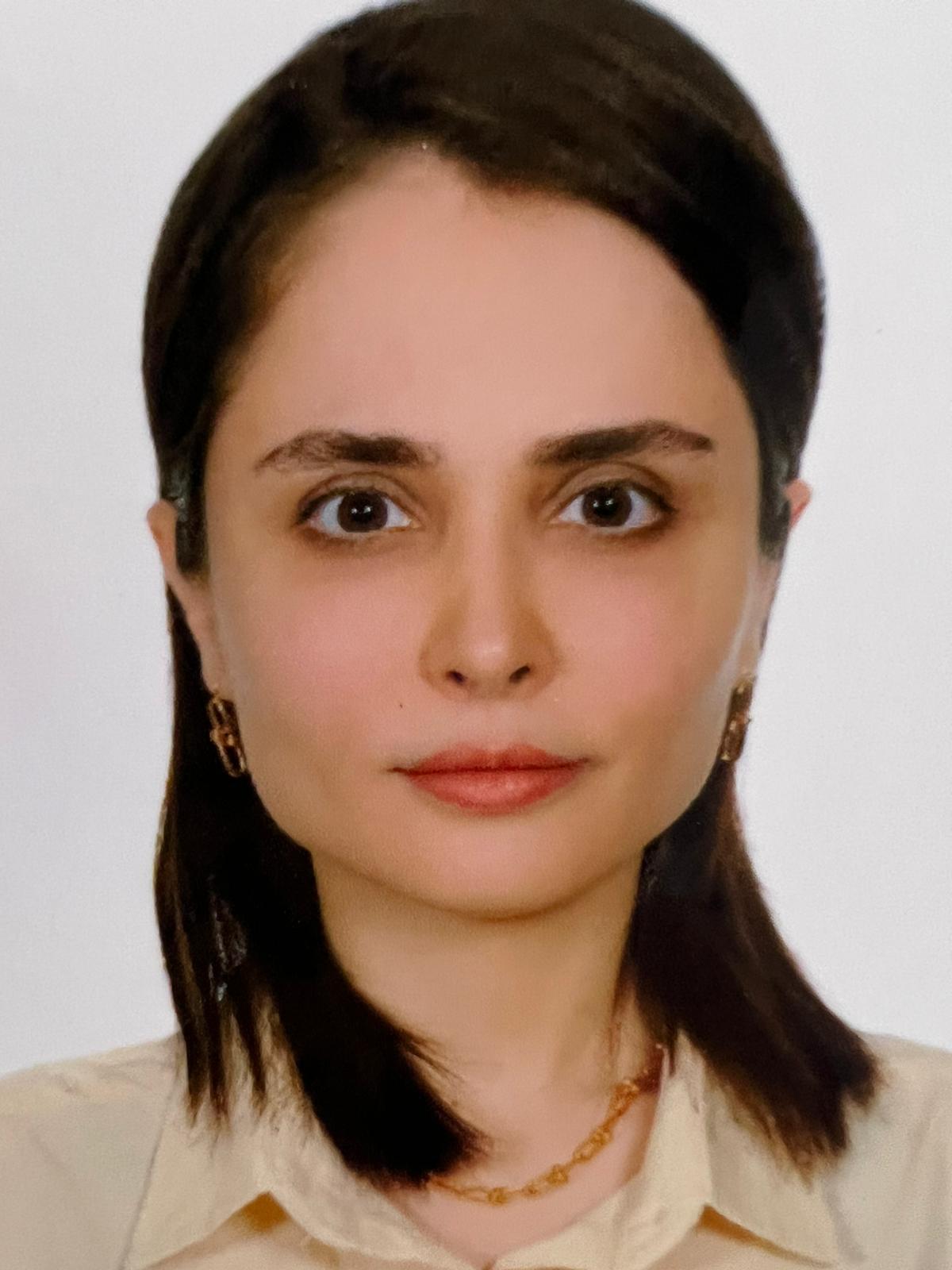}}]
{Samaneh Ghelichkhani} received a Master's degree in Advanced Computer Science from the University of Leeds, United Kingdom, and in Information Technology Engineering (Networking branch) from Islamic Azad University. Furthermore received a Bachelor’s degree in Information Technology Engineering from Islamic Azad University, Iran.  Her main research interest is in artificial intelligence including machine and deep learning, and networks.   \\
\end{IEEEbiography}

\vskip -2\baselineskip plus -1fil
\begin{IEEEbiography}
[{\includegraphics[width=0.9in,height=1.25in]{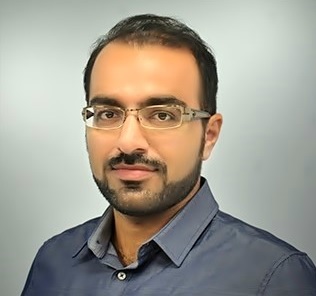}}]
{Ali Dehghantanha} is an academic-entrepreneur in cybersecurity, a Canada Research Chair in Cybersecurity and Threat Intelligence, and an Associate Professor in Cybersecurity at the University of Guelph, ON, Canada. He is the founding director of Canada Cyber Foundry, a research institute dedicated to advance research and training in cybersecurity - and the director and founder of the Master of Cybersecurity and Threat Intelligence program at the University of Guelph, Ontario, Canada.     \\
\end{IEEEbiography}

\end{document}